\documentclass[doublecol]{epl2}
\usepackage{arydshln}
\usepackage{amssymb,latexsym,mathrsfs}
\usepackage{amsfonts}
\usepackage{mathrsfs}
\usepackage{amsmath}
\usepackage{graphicx}
\usepackage{color}
\usepackage{ulem}

\def\bea{\begin{eqnarray}}
\def\eea{\end{eqnarray}}
\def\ba{\begin{array}}
\def\ea{\end{array}}

\def\la{\langle}
\def\ra{\rangle}

 \title{Self organised criticality in stochastic sandpiles: connection to Directed Percolation}
\shorttitle{SOC in stochastic sandpiles:  connection to DP}
 \author{Urna Basu \inst{1}  and P. K. Mohanty \inst{2,3}}
\shortauthor{Urna Basu and P. K. Mohanty}
\institute{ 
\inst{1}Instituut voor Theoretische Fysica, KU Leuven, 3000 Leuven, Belgium \\ 
\inst{2}CMP Division, Saha Institute of Nuclear Physics, 1/AF Bidhan Nagar, Kolkata-700064, India\\
\inst{3}Max-Planck-Institut f\"ur Physik komplexer Systeme, N\"othnitzer Strasse 38, 01187 Dresden, Germany}

\abstract{
 We introduce a stochastic sandpile model where finite drive and dissipation are coupled to the activity field. The absorbing phase transition here, as expected, belongs to the Directed Percolation (DP) universality class. We focus on the small drive and dissipation limit, {\it i.e.} the so called self organised critical (SOC) regime and show that the system exhibits 
a crossover from ordinary DP-scaling to a  dissipation controlled scaling 
which is  independent of underlying dynamics  or spatial dimension. The new  
scaling  regime   continues  all the way to the zero bulk drive limit  
suggesting   that the corresponding  SOC behaviour is only  DP, modified by  the dissipation-controlled scaling. We demonstrate this for continuous  and  discrete Manna Model  driven by noise and bulk dissipation.}
\begin{document}
\maketitle
\section{Introduction}
Sandpile models \cite{BTW, DR,Zhang, DD1990PRL,  Grassberger, Manna, Pietronero, Christensen, Dhar,PrussnerBook} 
show  scale free avalanche pattern  and are taken as  a  prototype models of  
self-organized criticality (SOC). In these models sand grains (particles or energy) 
are added  randomly to an empty lattice. Whenever the  number of grains in a  site 
crosses a predefined threshold value, it becomes unstable (active)  and relaxes by  toppling.  In a  toppling event particles or energy from each active site is redistributed   among the neighbours, which may further create  new  topplings. Such a cascade of  toppling events, commonly known as an avalanche,  continues in the 
system   until all sites become  stable (inactive); a new grain is added   then. 
The large avalanches usually hit the boundary where some energy is dissipated  out 
of the system.  The interplay of  the  slow driving, fast relaxation,  and dissipation 
at the boundaries brings  in a self-organized critical  state  without any 
fine-tuning of parameters.  It is well known that  critical behaviour  of sandpile models with stochastic 
toppling rules differ from those     having  of deterministic  dynamics and    
form a generic universality class, namely  Manna  class \cite{Manna}.

It was argued by Dickman  and co-workers~\cite{Dickman},   and supported by 
several other works \cite{FES_SOC}, that  the  critical behaviour of SOC can be understood as an ordinary 
absorbing phase transition (APT) in  a fixed energy sandpile (FES). The slow drive and boundary dissipation  
in SOC ensure that density gets adjusted to the critical value.
Since the most robust  universality class of absorbing state phase transition  
is DP   one naturally asks   whether  self organised criticality of 
stochastic sandpile models   are  in any way connected to DP. This doubt is bolstered by the fact that the exponents of Manna  class are not very different from DP. 
Several attempts have been made over last  decades to understand this riddle\cite{Tadic_DD, DD_PK, MunozETC}. In fact, both stochastic and deterministic   sandpile models flow  to DP when  perturbed\cite{DD_PK}.   It was also suggested recently 
that  the ordinary  critical behaviour  of fixed energy stochastic  sandpiles belongs 
to  DP\cite{nomanna}, though this issue is still being debated \cite{Lee-dickman}. All these works raise a possibility  that  observed self organised criticality in stochastic sandpile models are also related to DP. 

\begin{figure}
 \centering
 \includegraphics[width=4.4 cm]{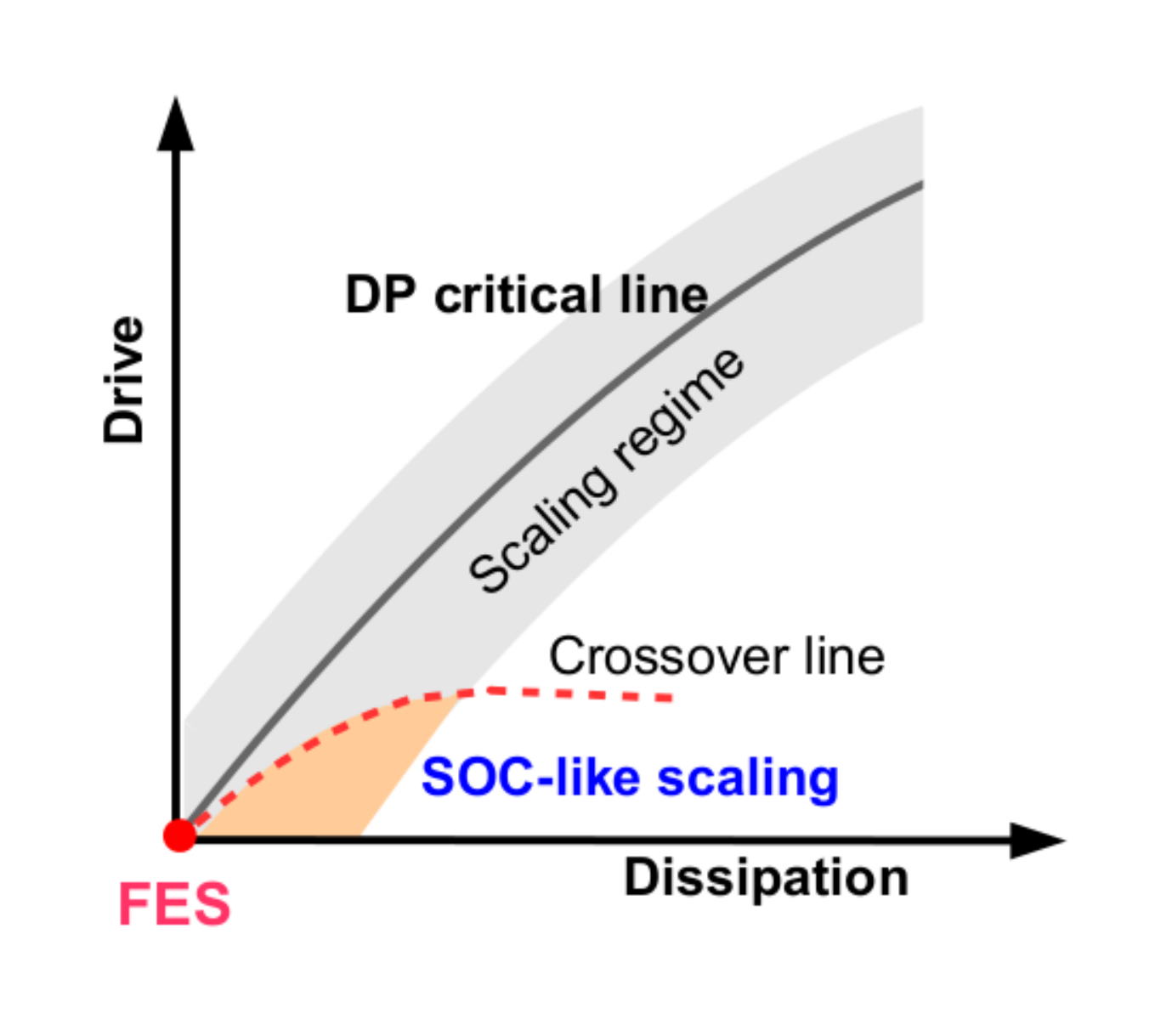}\includegraphics[width=4. cm]{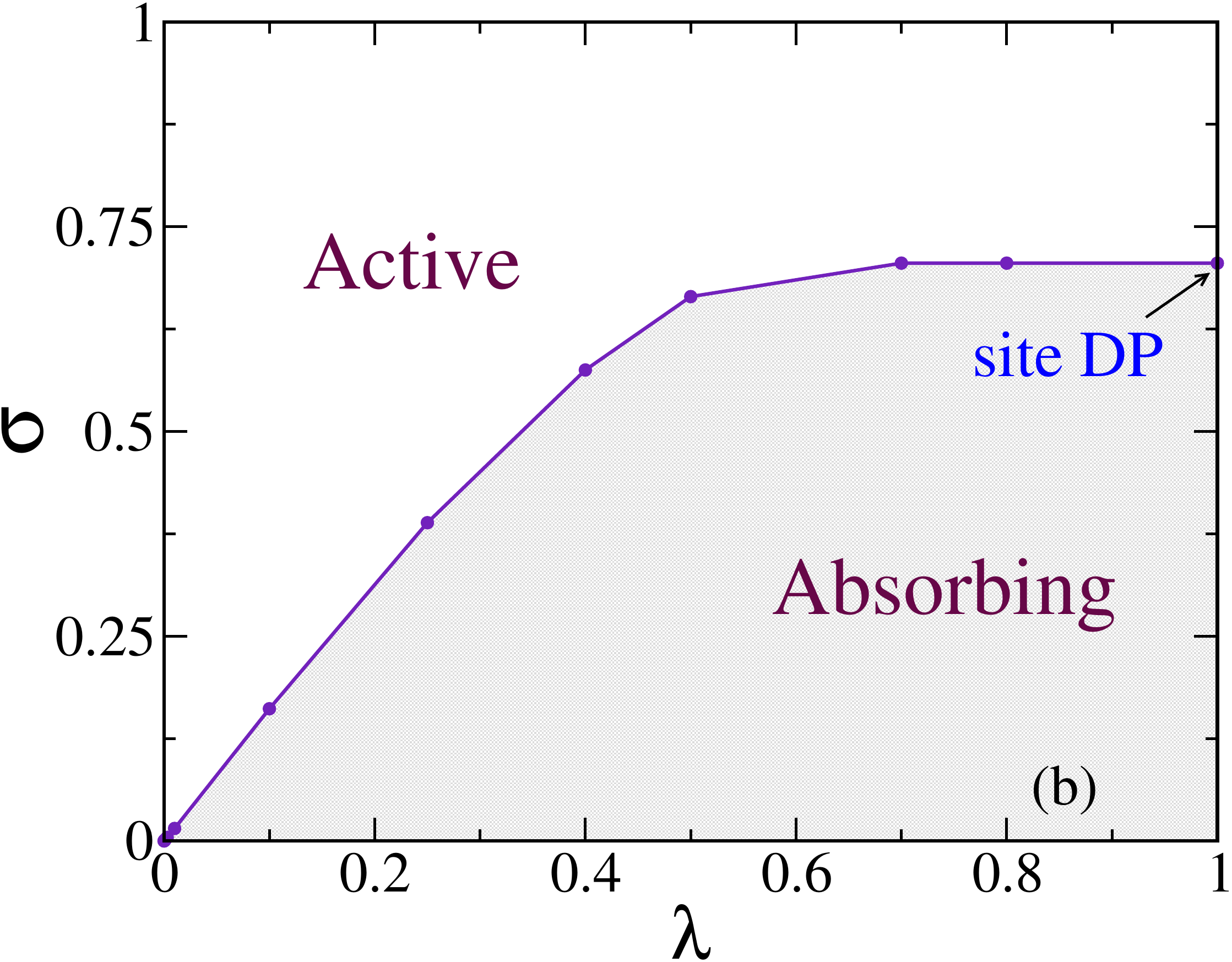}
 \vspace*{-0.3 cm}
 \caption{(a) Schematic phase diagram: Grey shaded area represents the scaling regime  around DP critical line.  
 In the small drive-dissipation regime, the cross-over line (dashed)  separates ordinary DP and 
 SOC-like scaling (orange shaded area). Ordinary DP scaling disappears in the zero  drive limit. 
 (b) Actual phase diagram for  driven dissipative Manna model in $1d$.  }
 \label{fig:schematic}
\end{figure}

 In this Letter we  attempt to explore this possibility   and bridge the gap between DP and SOC. Conventionally  self organised sandpile models
 are studied  with dissipation only at the boundaries. Another equivalent approach,
 where dissipation is incorporated in  the bulk of a closed system  \cite{DD_PK, bulk_dissipation}, has  certain advantages;  
it avoids difficulties like non-zero  particle current from the bulk towards  the boundary \cite {soc_current,Hwa}, inhomogeneous correlated height profiles \cite{Tadic_DD}  and  other unusual boundary   effects \cite{MunozETC}. 
Here  we choose to work with bulk dissipation (parametrized  by $\lambda$) and 
introduce additional finite drive $\sigma$   coupled to  the activity in a way 
that  the dynamics of the driven dissipative sandpile model reduces  to  SOC  
in $\sigma \to 0$ limit  and site-DP when  $\lambda \to 1.$
We find that in the small drive-dissipation limit,     
a  new scaling regime emerges  in the sub-critical  
phase when one moves away from the   critical  $\lambda_c$ $-$ observables which
ordinarily scale as  $(\lambda-\lambda_c)^a$  crosses over  to 
$(\lambda-\lambda_c)^{\tilde a}.$  We argue that the new  exponent 
$\tilde a$ can be expressed   in terms of known DP-exponents $a$ and 
$\gamma$ as  
\begin{equation}
 \tilde a = a/\gamma.
\end{equation}
This  new scaling regime    persists   all the way  down  to $\sigma= 0$ line suggesting, first, 
  what is commonly  know  as bulk dissipative SOC  (for $\sigma= 0$ and small $\lambda$)  
is nothing  but   DP with modified scaling, and  secondly,  a SOC-like behaviour can   
also be observed  in systems with finite drive and dissipation,  both local. A schematic representation of this scenario  
is presented in Fig. \ref{fig:schematic} (a). We use numerical simulation to verify this picture  for 
continuous Manna Model in  one dimension ($1d$)  and $2d$, and  discrete  Manna Model in $1d$.
     
\section{ Driven dissipative continuous  Manna Model}

Driven dissipative continuous   Manna  Model   can be defined  on  a   general  graph as follows.
 Each site  $\mathbf{R}$  on the graph has a continuous variable $E_\mathbf{R},$ called energy, associated with it; sites with $E_\mathbf{R} \ge 1$ are declared active.  At any given instant, let 
 $S_a$  be the set  of active sites  and  $S_n$,  the set of neighbour of the 
 active sites (which may or may not be active). The dynamics proceeds as a three step parallel  update: \\
{\bf I.} {\it Dissipation:} All sites  belonging to ${\cal S}_a \cup {\cal S}_n,$   $i.e.$ the sites which are active themselves or have at least one active neighbour,   dissipate  $\lambda$ fraction of their energies,
\begin{eqnarray} 
E_\mathbf{R}  \to  (1-\lambda) E_\mathbf{R}~~~~ \forall ~  \mathbf{R} \in {\cal S}_a \cup {\cal S}_n. \label{eq:I}  
\end{eqnarray}
{\bf II.} {\it Distribution:} All active sites  distribute their remaining energy randomly among 
the neighbours, {\it i.e.} for all $\mathbf{R} \in{\cal S}_a,$  if $N_ \mathbf{R}$
 is the set  containing  neighbours of $\mathbf{R}$
\begin{eqnarray}
E_{\mathbf{R'}}  &\to& E_{\mathbf{R'}}   + r_\mathbf{R'} E_ \mathbf{R} ~~~~ \forall ~ \mathbf{R'}  \in N_ \mathbf{R}\cr
{\rm and} ~~ E_{\mathbf{R}}&\to& 0  ~~~~~~~~~~~~~~~~~~~~\forall ~ \mathbf{R}\in{\cal S}_a,
  \label{eq:II}  
\end{eqnarray}
where 
$\{  r_{\mathbf{R'}} \in (0,1)\}$   are random numbers satisfying 
$$\sum_{\mathbf{R'} \in N_ {\mathbf{R}} }r_\mathbf{R'} =1.$$
{\bf III.} {\it Drive:} Finally, the drive is added with probability $\sigma$ independently to all the sites  belonging to ${\cal S}_n$ ($i.e.$ the receiving sites) in the form of,
\bea
   E_{\mathbf{R}} \to E_{\mathbf{R}} + 1 \;\;\;\; \text{with probability $\sigma.$}\label{eq:III}
\eea 
\\
Note that, in this dynamics, energy is added to or dissipated from the system  only when 
it is active, ensuring  that  absorbing  configurations  are   not spontaneously 
activated  by noise. 

Some of the limiting cases  of  this dynamics  are of special interest. Without any drive or dissipation $\sigma=0=\lambda,$ 
this model   maps  to  the  conserved continuous Manna model (CCMM) in $d$-dimension \cite{nomanna};  the  conserved density   
needs   to be tuned   to locate  the absorbing  phase transition  in this fixed energy   sandpile model. 
On the other hand, when $\lambda=1$ the active site surely becomes inactive after each update, and each of the sites which have at least one active neighbour  gets activated themselves with probability $\sigma.$ This is the dynamics of site directed percolation; thus  for $\lambda=1$ the present model  would show an absorbing state transition at \cite{dp}
\begin{equation}
\sigma_c^{DP}=\left\{\begin{array}{ll}
                       0.705489~~~  & d=1\cr 
                       0.34457~~~  &  d=2 {\rm ~square ~lattice }
                      \end{array}
 \right.\label{eq:sigma_c}
\end{equation}

For non-zero noise and 
dissipation,  the parameters $\lambda$ and $\sigma$ control the average energy 
of the system. However the absorbing configurations  of  this model    are no 
different from  those of  CCMM  since   the additional  dynamics {\bf I.} and 
{\bf III.} can not be executed  on inactive states. For any  given $\lambda,$ the system 
is expected to fall into an absorbing configuration when $\sigma$ is decreased below a
critical threshold $\sigma_c(\lambda).$ 
Since  the  model satisfies all the criteria of the DP conjecture\cite{dp_conjecture}, one naturally 
expects that the critical behaviour along the  critical line $\sigma_c(\lambda),$   which  
includes   the   site-DP critical point  $(\lambda=1,\sigma =\sigma^{DP}_c),$ would belong 
to the DP universality class.  

Let us consider the $d=1$ case in details. On a   one dimensional  periodic lattice with $L$ 
sites $i = 1, 2, \dots , L$, 
each having a  continuous  variable called  energy $E_i,$ the  three step parallel 
dynamics  reads as follows.
\noindent{\bf I.}  All sites  belonging to ${\cal S}_a \cup {\cal S}_n,$ dissipate  $\lambda$ fraction of their energies $E_i  \to  (1-\lambda) E_i$, 
and then  {\bf II.} the active sites distribute their remaining energy randomly among the two neighbours, i.e.  
$ E_{i \pm1}  \to  E_{i \pm1}  + [\frac 1 2  \pm (r_i - \frac 1 2)]  E_i$   and $E_i \to 0.$  Here $r_i$ is a random number  distributed uniformly in  $(0,1).$
 And finally, {\bf III.}  all the sites $i \in{\cal S}_n,$ 
are activated    by adding unit energy  independently and randomly  with probability $\sigma,$  i.e., $E_i   \xrightarrow{\sigma}  E_i  + 1.$

We have studied absorbing state phase transition here  for a set  of  values of
$\lambda$  taking $\sigma$ as the  tuning parameter $\sigma$ and verified 
explicitly that the  whole critical line $\sigma_c(\lambda),$  shown in  Fig. \ref{fig:schematic}(b)
belong to the DP universality class.  For details of this study see the supplementary  text \cite{Supple}.

\begin{figure}[t]
 \centering
 \includegraphics[width=7 cm]{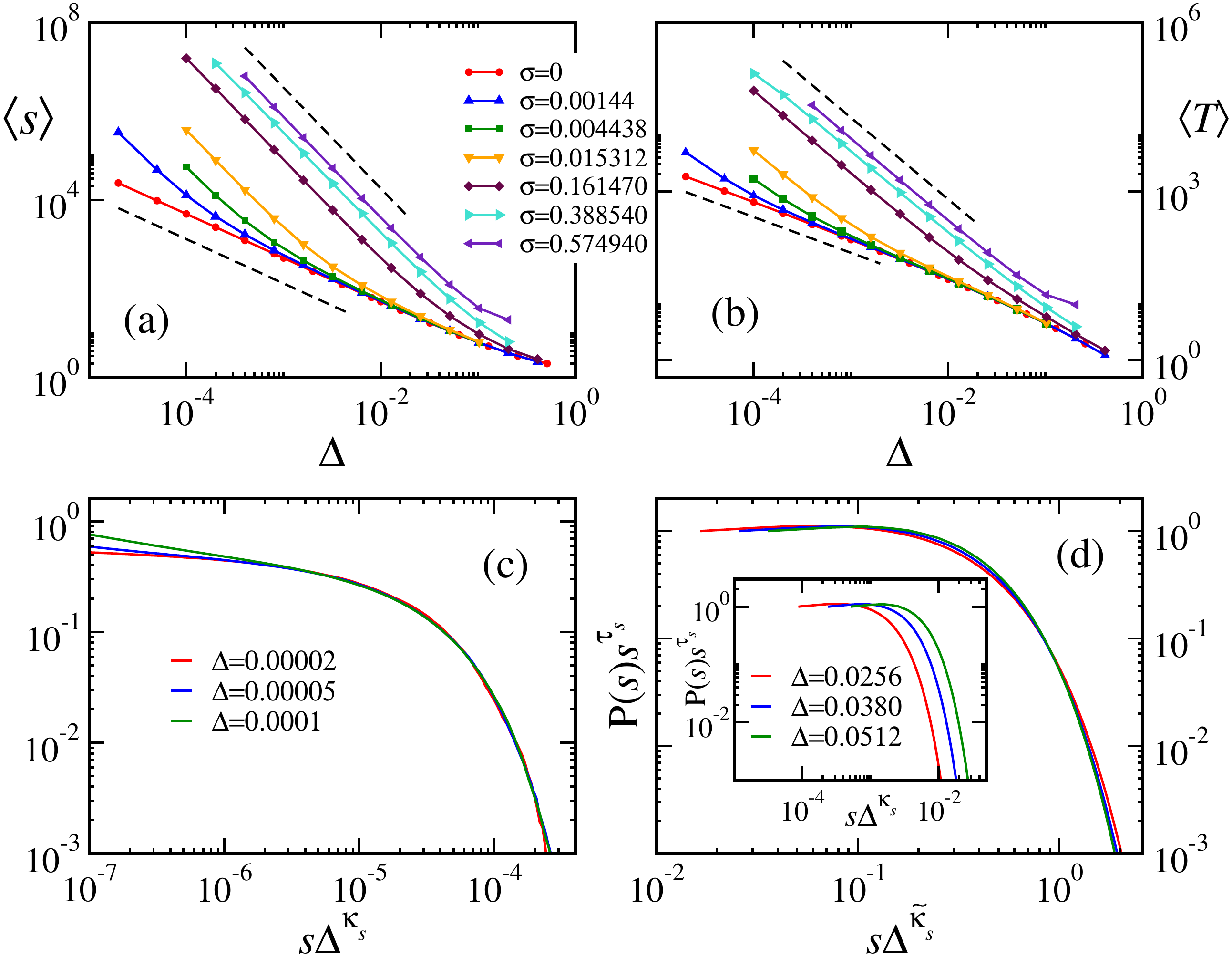} \\
\caption{ (a) The average   cluster size $\la s \ra$ and (b)  average lifetime $\la T\ra$ versus $\Delta=\lambda -\lambda_c$ for different values of $\sigma.$ 
(c) and (d)  shows scaling collapse of $P(s)$ following \eqref{eq:Ps}: 
  Curves corresponding to $\Delta= \lambda-\lambda_c= (2,5,10) \times 10^{-5}$  in (c) could be collapsed  with   
  DP value $\kappa_s=\gamma/(2-\tau_s)=2.55$  whereas the same  for   $\Delta= 0.0256,0.0380,0.0512$  in (d) are collapsed  
  with  the modified exponent  $\tilde \kappa_s=1/(2-\tau_s)=1.12.$  Here,  $L=10^4, \sigma=0.004438$ (corresponding  
  critical point  $\lambda_c=0.003$)  and the statistical  averaging  is  done  over $10^5$ to $10^7$ 
  independent clusters.}
\label{fig:seed}
\end{figure}

Our main aim is to study the  small drive-dissipation 
limit of this dynamics and to relate the critical behaviour  to SOC. One way   is to
generate clusters from a single seed in the sub-critical regime of the APT  and   ask  if 
their   statistics close to the critical point  relates  to  that of SOC \cite{scaling}.
To this end,  starting from a fully active state first the system is allowed to relax;
absorbing configurations are then activated  by   generating a  seed 
at a randomly chosen site by 
adding one unit of energy \cite{UnlikeFES}. This  {\it seed-simulation  process} is 
repeated  to  obtain statistics of the  clusters generated.
For any fixed $\sigma$   the clusters are expected to be characterised by   DP-critical exponents 
and scaling functions
\bea
P(s) \sim s^{-\tau_s} f(s\Delta^{\kappa_s}) ;~~P(T) \sim T^{-\tau_t} g(T \Delta^{\kappa_t}),\label{eq:Ps}
\eea
Here $\Delta \equiv \lambda - \lambda_c,$ and  $s$, $T$  denote the size and  lifetime of clusters.
Consequently their averages diverge as $\la s\ra \sim  \Delta^{-\gamma}$ and $\la T \ra \sim  \Delta^{-\tau}$ near the critical point with $\gamma = \kappa_s (2- \tau_s)$ and $\tau = \kappa_t (2- \tau_t)$ (see Table \ref{tab:exp}).
However,  the average energy added  and dissipated  per cluster 
must balance to maintain a  stationary  state;  this puts an additional  constraint \cite{delta} on  $ \la s\ra,$   effective  primarily
in the  small drive regime, and prompts 
\bea
\la s\ra \sim \Delta^{-1}.\label{eq:soc}
\eea 
This opens  up a possibility that $\la s \ra$ might show a different scaling for small drive $\sigma.$

 \begin{figure}[t]
 \centering
 \includegraphics[width=7 cm]{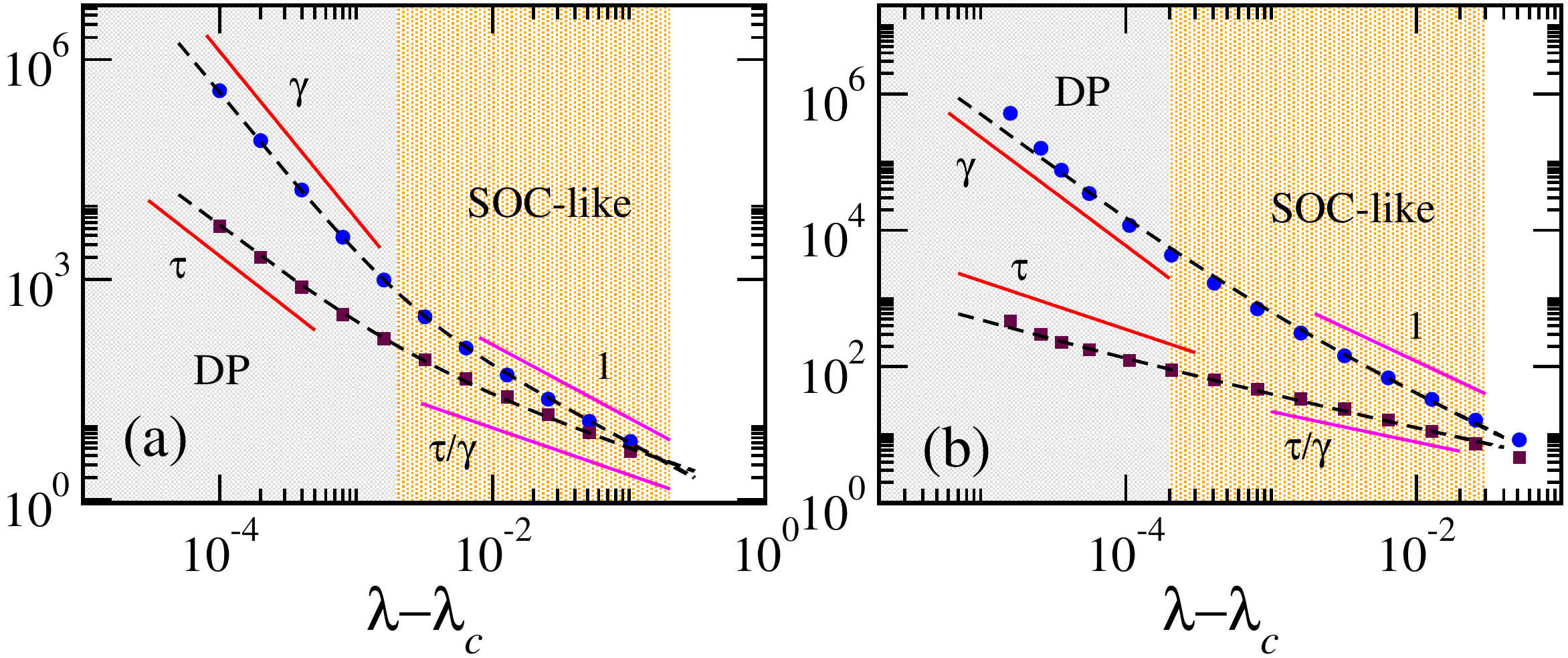}
 \caption{The DP and SOC-like scaling regimes: (a) $1d$ driven dissipative continuous Manna model. The data used for illustration corresponds to $\sigma=0.015312$ for which $\lambda_c=0.01.$ The dashed lines are best fit curves obtained following Eq. (\ref{eq:sav_tav}). The solid lines are for guidance. (b) The same for $2^{10} \times 2^{10}$ square lattice,   when 
 $\sigma=0.01$, $\lambda_c=0.011974.$  Here, statistical averaging is doen over $10^4$ to $10^6$ clusters.}
 \label{fig:dp_soc_region}
\end{figure}

Figure \ref{fig:seed}(a) shows plots of  $\la s \ra$ as a function of $\Delta \equiv \lambda - \lambda_c $ for different values of
$\sigma$ including $0.$   For large $\sigma$ the average cluster size  diverges as 
$\la s \ra \sim \Delta^{-\gamma}$ with $\gamma=2.277$ as expected for DP. 
However, a new scaling regime emerges as $\sigma$ is decreased; $\la s \ra$ shows a crossover from the DP behaviour to $\la s \ra \sim \Delta^{-\tilde \gamma}$ with $\tilde \gamma=1 $ as $\lambda$ is increased further away from the corresponding critical point $\lambda_c(\sigma)$.  We must emphasize that 
this   crossover is {\it not}  an artefact of  long relaxation time  or small system size. If that were
the case, unusual scaling would rather appear   closer  to  the critical point  as opposed to  what we 
see here, {\it i.e.} 
the  DP critical behaviour prevails   near  the critical line. This endorses  the fact  that the cluster 
statistics is obtained correctly - the system is fully relaxed  and the results do not suffer from  finite size effects. 

The crossover  starts at  smaller $\lambda$ as noise   strength $\sigma$ is decreased; indeed for $\sigma=0$ (lowest red curve)  the DP regime completely disappears, and we only see  
\begin{equation}                                                                                                                                                                                                                                                                                                                                                                                                                                                                  
 \la s \ra \sim \lambda^{-1}.    \label{eq:laminv}                                                                                                                                                                                                                                                                                                                                                                                                                                                  
\end{equation}

%
\begin{largetable}
 \caption{  Directed percolation exponents along with the modified  values  in SOC-like  scaling.}\label{tab:exp}
  \begin{tabular}{|c|cccccc|cccc|}
  \hline
  & $\tau_s$  &$\gamma$  & $\kappa_s$ &$ \tau_t $ &$   \tau$  & $\kappa_t$ & $  \tilde \gamma $  &$\tilde \kappa_s$ & 
  $\tilde \tau$ & $\tilde \kappa_t$\\   \hline
  $1d$& 1.108 & 2.277 & 2.553 & 1.159 & 1.45  & 1.724 & 1 & 1.121 & 0.636 & 0.757 \\
    \hline  
    $2d$& 1.267 & 1.594 & 2.174 & 1.457 & 0.712 & 1.295 & 1 & 1.364 & 0.447  & 0.812 \\
  \hline
  \end{tabular}
  \vspace*{-.5 cm}
  \end{largetable}

 This dissipation  controlled behaviour is characteristic to bulk dissipative SOC   models  \cite{DD_PK,PrussnerBook}. 
 Indeed  the $\sigma=0$  line   is the SOC-limit in this model as will   be discussed later  in this  article.
 Although dissipation controlled avalanches  are  familiar in  
SOC (rather  necessary to maintain a self-critical state) possibility of  
their presence  and effect in  ordinary   absorbing transition  is  explored in this work. 
What it   brings in  here  is a  crossover from ordinary DP-critical  
behaviour $\la s \ra \sim \Delta^{-\gamma}$  to  $\Delta^{-1}.$   The immediate question is then whether it affects other 
critical exponents.  Since the underlying   universality  is still DP (for any non-zero $\sigma$)
it is natural to expect that  other exponents  would be  modified  in a way that DP-signature   
is retained. We put forward a conjecture that 
Eq. (\ref{eq:soc}) prompts $\Delta \to \Delta^{1/\gamma}$ and  observables  which   ordinarily scale as $\Delta^a$   
would crossover to $\Delta^{\tilde a}$  with  $\tilde a= a/\gamma.$ 
Accordingly in the new scaling regime, which we refer to as `SOC-like' regime, 
the    cluster statistics  would  then be given by 
\bea
P(s) \sim s^{-\tau_s} f(s\Delta^{\tilde\kappa_s}) ;~~P(T) \sim T^{- \tau_t} g(T \Delta^{\tilde \kappa_t}),\label{eq:Ps_tld}
\eea
with  $\tilde \kappa_{s,t} = \kappa_{s,t}/\gamma$.   Consequently    $\la s \ra = \Delta^{-\tilde \gamma}$  and 
$\la T \ra=\Delta^{-\tilde \tau}$  with  $\tilde \gamma= \tilde\kappa_s (2- \tau_s)=1$ and  $\tilde \tau= 
\tilde\kappa_t (2- \tau_t)=\tau/\gamma;$ see Table \ref{tab:exp} for the numerical values.   

 To verify this proposition  we   measure  $\la T \ra$  and  $P(s).$  Figure \ref{fig:seed}(b)  shows  $\la T \ra$ as a function of $\Delta$ for different values of $\sigma.$  Clearly, DP behaviour prevails near the critical point whereas 
the exponent that dictates  $\la T \ra$ further away is nothing but $\tilde \tau  =\tau/\gamma.$
Change  in   the functional form of  $P(s),$  from Eq. (\ref{eq:Ps}) to Eq. (\ref{eq:Ps_tld})
can be captured from the data collapse  of   $P(s) s^{\tau_s}$ as a function of  $s \Delta^{\kappa_s}.$ 
As seen in   Fig. \ref{fig:seed}(c), use DP  exponent  $\kappa_s$  results in  a perfect data collapse  
for  small values of $\Delta,$  but  fails for   relatively  large $\Delta$ (inset of Fig. \ref{fig:seed}(d)). 
$P(s)$ data for larger $\Delta$  could be collapsed with modified  DP exponent $\tilde \kappa_s=\kappa_s/\gamma.$

At this point,  we  arrive at the following picture - the sub-critical scaling regime of the driven dissipative 
Manna model in $1d$  is divided into two regions in the small drive-dissipation limit, 
ordinary DP scaling  near the critical line  crossing over to an  emerging SOC-like scaling   as one moves away.
This is depicted in  Fig. \ref{fig:schematic} (a)  with a  schematic crossover  line  that   
separates  DP  and  SOC-like scaling; the actual phase diagram in  Fig. \ref{fig:schematic} (b) 
shows the critical line.

Let us look at the generality of this scenario.
Eq. (\ref{eq:laminv}), which originates from a generic  energy balance condition in the stationary state \cite{delta}, 
is  expected  to  hold  in  other stochastic sandpile models,  in one and higher dimensions;  
the crossover  from ordinary DP to   SOC-like  scaling in the sub-critical regime can be viewed 
as  a generic multi-scale behaviour, 
\bea
\la s \ra = A_s \Delta^{-\gamma} + B_s \Delta^{-\tilde \gamma} ~;~
\la T \ra = A_t\Delta^{-\tau} + B_t \Delta^{-\tilde \tau} \label{eq:sav_tav}
\eea
where $\sigma$ dependent coefficients $A_{s,t},B_{s,t}$ determine the crossover scale.
 Of course, the DP exponents  $\gamma$ and  $\tau$   
depend on  spatial dimension $d$,  but they would still rescale  as $\tilde \gamma=1$, 
$\tilde \tau = \tau /\gamma.$ In Fig. \ref{fig:dp_soc_region}(a) and (b)  we  verify the 
same for continuous Manna model in  $1d$ and $2d$  respectively. The dashed lines there   
are the  best  fit of the data points  according to  Eq. \eqref{eq:sav_tav},   with 
exponents in  Table \ref{tab:exp}.  

Now we turn our  attention to the $\sigma=0$ line.  Here the   unit energy added 
to  create an active seed initiates a  cluster  which runs  until all 
the  sites become inactive;   there is no energy  input during the propagation. 
This is exactly  how   avalanches are created   and propagated in the corresponding 
SOC models. The  average energy dissipated per cluster  is proportional to 
$\lambda \la s \ra,$  which must balance the  unit energy added initially, 
leading  to  Eq. (\ref{eq:laminv})  under  stationary condition. 
This  condition, as we have already mentioned,  is common to {\it all} 
bulk dissipative  SOC models. It  also has a well known analogue
in context  of boundary dissipative sandpiles.  There $\la s \ra \sim  L^2$  \cite{DD1990PRL}  
independent of the  dynamics \cite{Biham} and spatial dimension \cite{Grassberger}. 
In boundary dissipative  SOC models the   slow dissipation required to reach self critical state
is naturally achieved     by taking  $L\to \infty.$
In contrast, models with  bulk  dissipation are  studied in thermodynamically large 
systems and criticality is reached  in the limit $\lambda \to  0.$ 

\begin{figure}[t]
 \centering
 \includegraphics[width=7 cm]{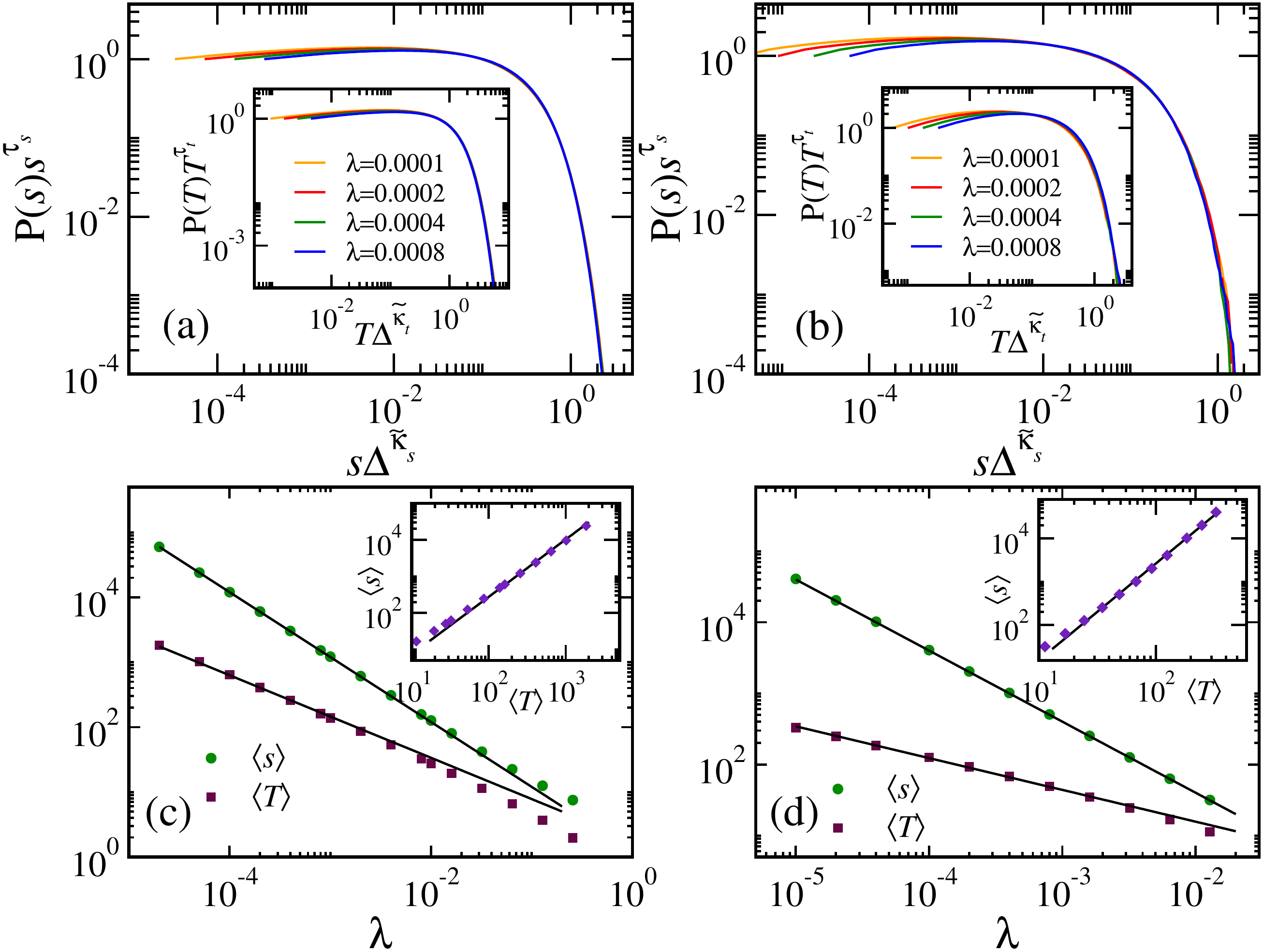} 
\caption{The SOC limit $\sigma=0$ (a)$1d$ continuous model : Data collapse for $P(s)$ according to \eqref{eq:Ps_tld} with $\tilde\kappa_s=1.121$ for different values of $\lambda.$ The inset shows the same for $P(T)$ with $\tilde\kappa_t=0.757.$ (b) The same as (a) for $2d$ continuous model. Here $\tilde\kappa_s=1.364,\tilde\kappa_t=0.812.$ 
(c) Plot of $\la s \ra$  and $\la T \ra$ with $\lambda.$   Corresponding exponents are  $\tilde \gamma=1$ and $\tilde \tau=.636$  (solid lines). 
The inset shows $\la s\ra \sim \la T \ra ^{\gamma/\tau}.$  (b) and (d)  are  same as (a) and (c),  but  
for  $2d$ continuous model  with   exponents  in Table \ref{tab:exp}.}
\label{fig:soc}
\end{figure}

Next  we  explore  whether  the modified  DP scaling  seen for non-zero but small $\sigma$   persists  
up to the SOC line  $\sigma=0.$ If this  scenario continues  all the way to  $\sigma=0,$   one must observe that the  avalanche statistics  there obey  Eq. (\ref{eq:Ps_tld})   with  the modified exponents.
In Fig. \ref{fig:soc} we have verified this both for $d=1,2.$  
 The data collapse according to Eq.  \eqref{eq:Ps_tld} for $P(s)$ and $P(T)$ are shown  in Fig. \ref{fig:soc}(a)  
 and its  inset respectively. Figure \ref{fig:soc}(c) shows plots of $\la s\ra$ and $\la T \ra$ 
 which  clearly   agree with  modified  DP-exponents  $\tilde \gamma=1$ and $\tilde \tau = \tau/\gamma.$ 
 This particular modification    does not  affect  the  DP-scaling form  $\la s\ra \sim \la T\ra ^{\gamma/\tau}.$  
 Indeed, the plot of $\la s\ra$  vs. $\la T\ra$   in inset of   Fig. \ref{fig:soc}(c)  shows that 
 the DP-exponent  $\gamma/\tau$  is retained  even in the SOC. 
 The same  scenario   also holds in  higher dimensions -Fig. \ref{fig:soc}(b) and (d) demonstrate it for  the 
 driven dissipative  Manna sandpile model in $2d.$

 These results  encourage us to propose    that 
 the critical behaviour of bulk dissipative SOC is  only   DP  with rescaled exponents. 
 To provide additional evidence next we study the stochastic sandpile models  with discrete variables \cite{Manna} by adding  finite drive and diffusion. 

 \section{ Driven dissipative discrete Manna model}
 
The drive dissipation mechanism  can be  extended to   sandpile models    with discrete  variables. 
In context of  Manna model, we  consider a    $d$ dimensional cube  with  each site  $\mathbf{R}$ holding 
a {\it discrete} variable $n_\mathbf{R}$  called particle  number -sites  with  $n_\mathbf{R} \ge n_c,$  a predefined threshold  (usually $2$), 
hops  independently  to  a randomly chosen neighbouring site $\mathbf{R'}.$ 
Drive and dissipation  can be implemented as follows, \\
{\bf I.} {\it Dissipation :} All  the sites belonging to $S_a \cup S_n$ 
 attempt {\it independently} with probability $\lambda$ to   get vacated, {\it i.e.} 
 to throw out  all the  particles out of the system,   
\bea
n_ \mathbf{R} \to 0 ~~~\text{with probability $\lambda$}
\eea
{\bf II.} {\it Distribution :}  The active sites which did not dissipate distribute their  particles   to  the 
neighbouring sites; each particle independently moves to one   of the neighbours. \\
{\bf III.} {\it  Drive :} Finally, the sites belonging to $S_n$ $i.e$ all the neighbours 
of active sites are activated with probability $\sigma,$ by receiving $n_c$ 
particles 
\bea
n_\mathbf{R}  \to n_\mathbf{R} +n_c~~~\text{with probability $\sigma$}
\eea
Unlike the  continuous  case  (Eqs. \ref{eq:I}-\ref{eq:III}) here  $\lambda$  denotes the probability  with which 
a site decides to dissipate.

\begin{figure}[t]
  \centering
 \includegraphics[width=7 cm ]{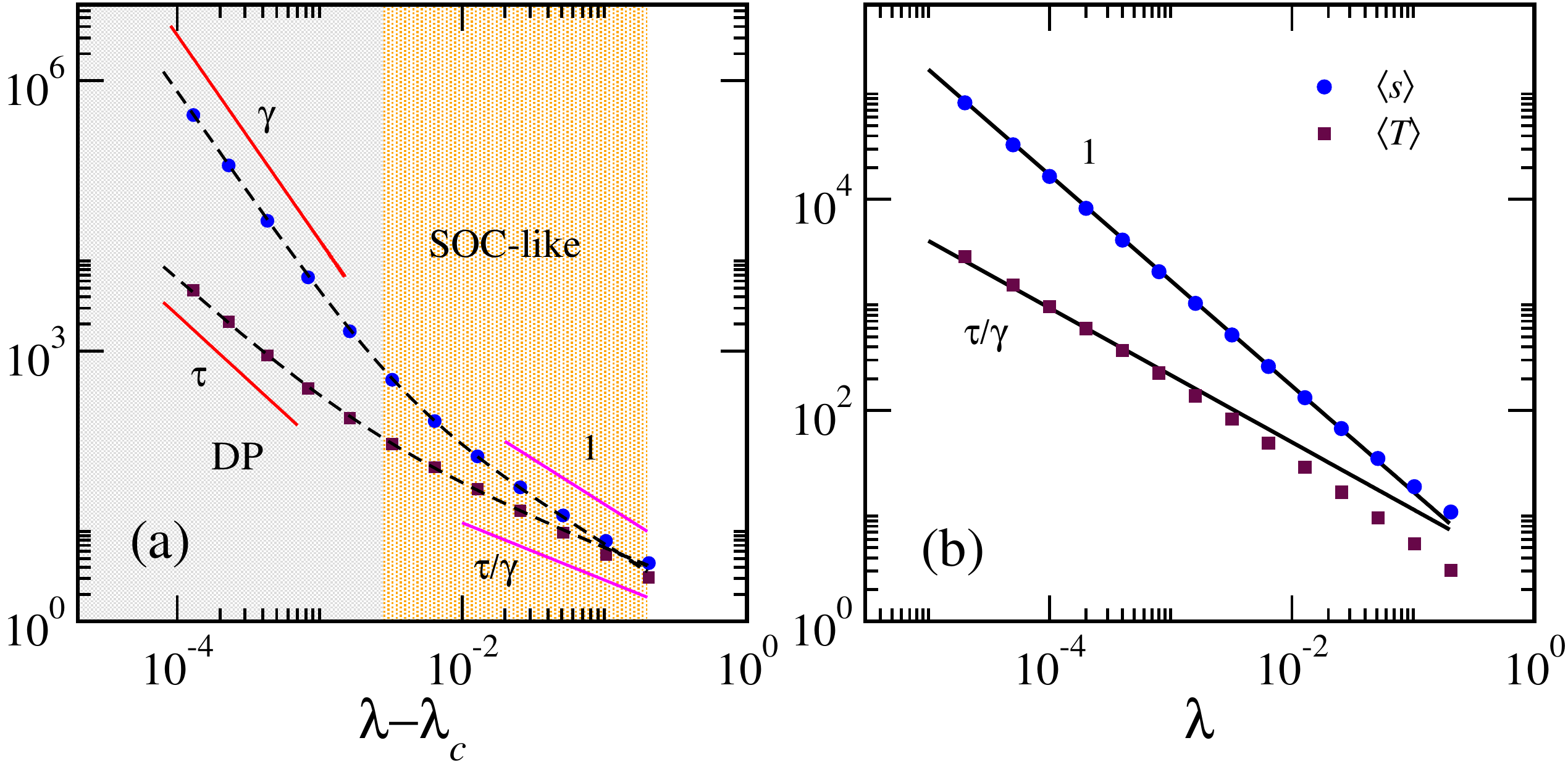}
 \caption {Discrete Manna model ($1d$) with drive and dissipation : (a) The crossover from DP to SOC like scaling for $\la s\ra$ and $\la T \ra$ against $\Delta=\lambda-\lambda_c.$ Here $\sigma=0.01$ and corresponding $\lambda_c=0.00757(1).$ The dashed lines  are best fit curves following Eq. \eqref{eq:sav_tav}. 
 (b) Plot of $\la s\ra$ and $\la T \ra$ for $\sigma=0.$ The solid lines correspond to the modified exponents $\tilde \gamma=1, \tilde \tau=0.636.$ Here $L=10^4$ and the data are averaged over $10^5$ or more ensembles.} \label{fig:dcmm}
\end{figure}

In the absence of drive and dissipation $\lambda=0=\sigma$ this dynamics only allows 
distribution of the particles of the active site and indeed is identical to the  well known 
fixed energy Manna model\cite{Manna}. When the conserved particle density $\rho=\frac 1 L \sum_i n_i$ is 
tuned beyond the critical value $\rho_c=0.89236$, this model undergoes an ordinary APT belonging to directed percolation universality class \cite{nomanna}. 
On the other hand, in the maximal dissipation limit $\lambda=1$ the dynamics once again becomes exactly that of site directed percolation with active sites infecting their neighbours with probability $\sigma.$ As expected, this shows 
an absorbing phase transition at $\sigma_c^{DP}=0.705489$ \cite{dp}.  
For any non-zero $\sigma < \sigma_c^{DP},$  the  discrete model 
undergoes a phase transition at a critical $\lambda_c < 1,$  belonging to 
DP-class (details are omitted here as this study closely follows  the one for the continuous version). 

In the small drive-dissipation limit,   $\la s\ra$ and $\la T \ra$   of the discrete model too show a crossover 
from DP to SOC-like  scaling, which  is described Fig. \ref{fig:dcmm}(a). 
The dashed lines   here  are  the best fit to  the data points  according to     Eq. \eqref{eq:sav_tav}.
Again for   $\sigma=0,$   as shown in   Fig. \ref{fig:dcmm}(b), the ordinary DP  feature is completely lost  
and one observes only SOC-like scaling.  Thus we conclude  that discrete version of the driven dissipative Manna model
in $1d$  also shows a crossover from DP to SOC-like  scaling, which   continues all the way  to 
the SOC line $\sigma=0$.

\section{Stochastic sandpile models with boundary dissipation}
Conventionally  self organised criticality  is  modeled  with boundary   dissipation   and  no 
additional drive ($i.e.$  $\sigma=0=\lambda$).  
Usually a particle  or unit energy is added  
to a  randomly chosen site  to  initiate an avalanche    which    propagates following  
a  conserving dynamics; avalanches which reach   the boundary  
can dissipate particles  at the boundary.   With increasing of system size, 
the  dissipation  rate decreases  and  accordingly  avalanche  size increases    
such that, on the average,   one  particle is   dissipated  per cluster.   Since the  conserving
bulk dynamics    can be described by a  diffusive process,  the   average    size of  avalanche  $\la s \ra$ 
is proportional   to the residence time  of a random walker (starting from a random site)  
on the lattice  with absorbing boundary,
\begin{equation} 
 \la s \ra  \sim L^2,  \label{eq:Lsq}
\end{equation}
where $L$ is the linear size of the lattice. 
Similarity between  \eqref{eq:Lsq} and  $\la s \ra  \sim \lambda^{-1}$   obtained for  bulk dissipative models (in Eq. \eqref{eq:soc}) 
is  that both  are derived from the  requirement of stationarity   and  hold  independent of type or spatial   
dimension of the  lattice.

\begin{table}
\begin{tabular}{ccc}
\hline 
Dissipation : & Bulk & Boundary\cr \hdashline
Critical  limit& $ \lambda\to 0$ & ${L\to \infty}$\cr  \hline%
 &  $\sim s^{-\tau_s} f(s\Delta^{\tilde\kappa_s})$ & $\sim s^{-\tau_s} f\left(s/L^{D_s}\right)$\cr 
$P(s)$ &   $\Rightarrow  \la s\ra \sim \lambda^{-\tilde \gamma}   $ &$ \Rightarrow  \la s\ra \sim L^{\mu_s} $  \cr 
& $\tilde \gamma =\kappa_s(2- \tau_s)$ &  $\mu_s = D_s(2- \tau_s)$ \cr\hline
&  $\sim T^{-\tau_t} g(s\Delta^{\tilde\kappa_t})$ & $\sim T^{-\tau_t} g\left(s/L^{D_t}\right)$\cr  
  $P(T)$ &     $\Rightarrow \la T\ra \sim \Delta^{-\tilde \tau}$ &  $\Rightarrow \la T\ra \sim L^{\mu_t} $\cr
 & $\tilde \tau= \kappa_t(2- \tau_t)$  & $\mu_t = D_t(2- \tau_t)$\cr 
\hline
Constraint &  $ \la s \ra \sim\lambda^{-1}$   &   $\la s \ra \sim L^2$ \cr\hline
\end{tabular}
 \caption{Connecting  bulk and boundary dissipative SOC.}\label{tab:lam_L}
\end{table}

 It was shown in Ref. \cite{Malcai} that   cluster statistics   of bulk dissipative    SOC   can me made equivalent  to  those  obtained   from boundary  dissipation  by  using  a system-size dependent dissipation parameter $\lambda$. Following this argument, we use an equivalence $\lambda \sim L^{-2}$    to   calculate  the  exponents   of 
SOC models with boundary dissipation (see  Table \ref{tab:compare}) from the exponents  obtained in this work. 
They can be expressed  in terms of DP-exponents   as
\begin{equation} 
  \mu_t=  2\tilde \tau = \frac  2 \gamma \tau    ~~ {\rm  and }~~  D_{s,t} = 2 \tilde \kappa_{s,t}= \frac  2 \gamma\kappa_{s,t}  \label{eq:Dst}
\end{equation}
Note that $\tau_{s,t}$ are not affected, they remain same as in ordinary DP.
  In  Table \ref{tab:compare}   we  compare the  recent numerical estimates  of  the exponents, 
  measured   in discrete  Manna model with boundary dissipation,  with Eq. \eqref{eq:Dst}.
  They  are  in  good agreement  - small discrepancies  could   come from   discreteness 
  (energy versus particle).  Study  of  continuous models, with boundary dissipation, is desirable.  
 \begin{table}
\begin{tabular}{lllll}\hline
 &$D_s$ & $\tau_s$ &$ D_t$ & $\tau_t$  \cr\hline
SOC($1d$)& 2.253(14) &1.112(6) &1.445(10)&1.18(2) \cr
Eq. (13)&2.242&1.108 & 1.514& 1.159\cr\hline
SOC($2d$)& 2.750(6) & 1.273(2)& 1.532(8)& 1.4896\cr
Eq. (13)&2.728&1.267 &1.624&1.457 \cr \hline
\end{tabular}
\caption{ Exponents obtained (from Ref. \cite{SOCExp})  for SOC models with boundary dissipation, 
are compared  with  what one  expects from  the modified-DP picture,  i.e. from  Eq.  (\ref{eq:Dst}).  } \label{tab:compare}
 \end{table}
 
%


\section{Conclusion}

 To summarize, in this article we study absorbing phase transitions in stochastic sandpile models 
 in presence of bulk drive and dissipation,  both coupled to activity. To facilitate the study of connection between SOC and DP, the stochastic sandpile model is designed in a way that in any dimension the dynamics reduces to site directed percolation and self organised criticality in two limiting cases.   In addition to the  generic DP critical behaviour,  in the slow drive dissipation regime  the system 
 shows a crossover in the  subcritical phase from ordinary DP to SOC-like  scaling.  We explain that 
 the exponents that characterise  the emergent scaling regime  are different, but can be  expressed in terms of DP-exponents. Moreover, this SOC-like scaling continues up to zero dissipation line (SOC  limit). Hence we argue that the critical   behaviour of bulk dissipative SOC  is only  DP,   modified by the dissipation control. We illustrate these phenomena explicitly for  continuous Manna model  in $1d$ and $2d$, and  its discrete version in $1d.$
These   results are not   restrictive to the  specific way   we  
introduce drive and dissipation here.    
What is important is that they must be coupled  to the activity 
field   so that  the  absorbing   configurations are not 
destroyed \cite{EPJB}. The specific  drive dissipation mechanism 
used here has an advantage   -it maps to  the site-DP  model 
for $\lambda=1.$ 

 Self organized  criticality in stochastic sandpile models,  conventionally studied   with  boundary dissipation,
is believed to belong to the Manna universality class.
Our results in context of    bulk dissipative SOC  suggest  what is ordinarily known as
Manna class  is  possibly  DP  with modified exponents. It would be interesting to explore  models 
with boundary dissipation    from   this  point of view.

  \acknowledgments{PKM  would   like to  acknowledge the support of CEFIPRA  under Project 4604-3.}

\end{document}


\title{Supplemental material for ``Self organised criticality in stochastic sandpiles: connection to Directed Percolation''} 
\author{Urna Basu$^1$ and P. K. Mohanty$^{2,3}$}

\affiliation{$^1$Instituut voor Theoretische Fysica, KU Leuven, 3000 Leuven, Belgium \\ 
$^2$CMP Division, Saha Institute of Nuclear Physics, 1/AF Bidhan Nagar, Kolkata-700064, India\\
$^3$Max-Planck-Institut f\"ur Physik komplexer Systeme, N\"othnitzer Str. 38, 01187 Dresden, Germany}

\begin{abstract}
In this supplemental text we provide more arguments and details about the numerical simulations  to
support our conjecture  presented in  Ref. \cite{main_text}  that the critical behaviour of driven dissipative sandpiles in the self-organised limit 
is nothing but directed percolation  with modified scaling.
\end{abstract}

\maketitle

\section{Absorbing Phase transition in driven dissipative Manna Model}

We study absorbing phase transition (APT) in driven dissipative stochastic sandpile models, particularly in the small drive-dissipation limit. 
The dissipation can be added naturally as a part of the relaxation mechanism of the active sites. But the important point is that  the bulk drive also  needs to be coupled to activity field  so that the  inactive sites  do  not get  activated spontaneously by  noise, destroying the transition altogether.  
 Note that in  usual {\clr  self organised critical (SOC)  models  of stochastic sandpiles  (in absence of bulk drive) \cite{Manna}  }
 particle  or energy is added only when the  system  is  inactive,   thereby  the system  either goes   
to another  inactive state  or   initiates  a  new avalanche. It is  
only in the latter case, the system becomes  active and experience  the actual dynamics of the model; 
in other words,   the operative drive  in SOC  is coupled to  the activity field. 
Coupling of drive  to activity  mimics this feature  of SOC.

We discuss a specific mechanism 
that  adds   drive  $\sigma$  and dissipation $\lambda$  to   stochastic sandpile models  on  a generic  graph, 
in a way that the model   reduces  to   site directed percolation (DP)   on    the same graph 
 for maximum dissipation  $\lambda=1.$   We show that    the   models  generically undergo a DP transition  and  the critical 
line,   starting from   the  site-DP   critical point      at $\lambda=1$, 
approaches $\lambda \to 0$  when   $\sigma = 0.$  In the  small drive-dissipation regime  
the system    shows a crossover   from DP  to SOC-like  scaling   in the  subcritical phase - 
 the  exponents    in the  SOC-like   scaling regime    are different, but can be  expressed in terms of 
 DP-exponents.    Moreover,  the usual  DP scaling  regime disappears  when $\sigma=0$   indicating 
 that   what we   usually know as bulk dissipative  SOC  (for  $\sigma=0$) 
is  only DP   with  modified  scaling behaviour.  
We  demonstrate this  for   continuous Manna Model in $d=1,2$  and for  discrete   Manna  
model in $d=1.$

\section{Driven dissipative Manna Model}

{\bf Continuous version:}  Driven dissipative continuous   Manna  model   can be defined  on  a   general  graph as follows.
 Each site  $\mathbf{R}$  on the graph has a continuous variable $E_\mathbf{R},$ called energy, associated with it; 
 sites with $E_\mathbf{R} \ge 1$ are declared active.  At any given instant, let 
 $S_a$  be the set  of active sites  and  $S_n$,  the set of neighbour of the 
 active sites (which may or may not be active). The dynamics proceeds as a three step parallel  update: \\
{\bf I.} {\it Dissipation:} All sites  belonging to ${\cal S}_a \cup {\cal S}_n,$   $i.e.$ the sites which are active themselves or have at least one active neighbour,   dissipate  $\lambda$ fraction of their energies,
\begin{eqnarray} 
E_\mathbf{R}  \to  (1-\lambda) E_\mathbf{R}~~~~ \forall ~  \mathbf{R} \in {\cal S}_a \cup {\cal S}_n. \label{eq:I}  
\end{eqnarray}
{\bf II.} {\it Distribution:} All active sites  distribute their remaining energy randomly among 
the neighbours, {\it i.e.} for all $\mathbf{R} \in{\cal S}_a,$  if $N_ \mathbf{R}$
 is set  containing  neighbours of $\mathbf{R}$
\begin{eqnarray}
E_{\mathbf{R'}}  &\to& E_{\mathbf{R'}}   + r_\mathbf{R'}  (1-\lambda)E_ \mathbf{R} ~~~~ \forall ~ \mathbf{R'}  \in N_ \mathbf{R}
  \label{eq:II}  
\end{eqnarray}
where 
$\{  r_{\mathbf{R'}} \in (0,1)\}$   are random numbers satisfying 
$$\sum_{\mathbf{R'} \in N_ {\mathbf{R}} }r_\mathbf{R'} =1.$$
{\bf III.} {\it Drive:} Finally, the drive is added with probability $\sigma$ independently to all the receiving sites  in the form of,
\bea
E_{\mathbf{R}} \to E_{\mathbf{R}} + 1 \;\;\;\; \text{with probability $\sigma.$}\label{eq:III}
\eea 
\\
{\bf Discrete version:} The drive dissipation mechanism  can be     extended to   sandpile models    with discrete  variables $n_\mathbf{R}$ at 
sites $\mathbf{R}$   on a  graph.
{\bf I.} {\it Dissipation :} All the the sites belonging to $S_a \cup S_n$ 
 attempt {\it independently} with probability $\lambda$ to   get vacated, {\it i.e.} 
 to throw out  all the  particles out of the system,   
\bea
n_ \mathbf{R} \to 0 ~~~\text{with probability $\lambda$}
\eea
{\bf II.} {\it Distribution :}  The active sites which did not dissipate distribute their  particles   to  the 
neighbouring sites; each particle independently moves to one   of the neighbours. \\
{\bf III.} {\it  Drive :} Finally, the sites belonging to $S_n$ $i.e$ all the neighbours 
of active sites are activated with probability $\sigma,$ by receiving $n_c$ 
particles 
\bea
n_\mathbf{R}  \to n_\mathbf{R} +n_c~~~\text{with probability $\sigma$}
\eea
The difference with  the continuous case  is that  in the dissipation part,  $\lambda$ denotes the probability 
 a site which decides to dissipate.

As discussed in the main text \cite{main_text}, the limiting case $\sigma=0=\lambda,$   of these models   are 
the  conserved Manna Models and   $\lambda=1$  case  maps to  site directed percolation. 
Since  the  model satisfies all the criteria of the DP conjecture \cite{dp_conjecture}, 
 we expect that the,   in generic values,  the absorbing  phase transition    obtained  by tuning  
 either or both    $\sigma$ and $\lambda$      belong to DP universality class. 
 Here   we verify this explicitly    for   continuous     Manna Model in  $1d$ for a  set of  
 $\lambda$  and tuning $\sigma.$  
 
{\clr  We report  the   details   for  $\lambda=0.1$;  this  would serve as an   illustrative example of how we determine the critical point and   
 some  of the exponents.    The   other values of  $\lambda$ that we studied  are  listed in   Table  \ref{tab:1dccmm}.}
\begin{figure}
 \centering
 \includegraphics[width=8 cm ]{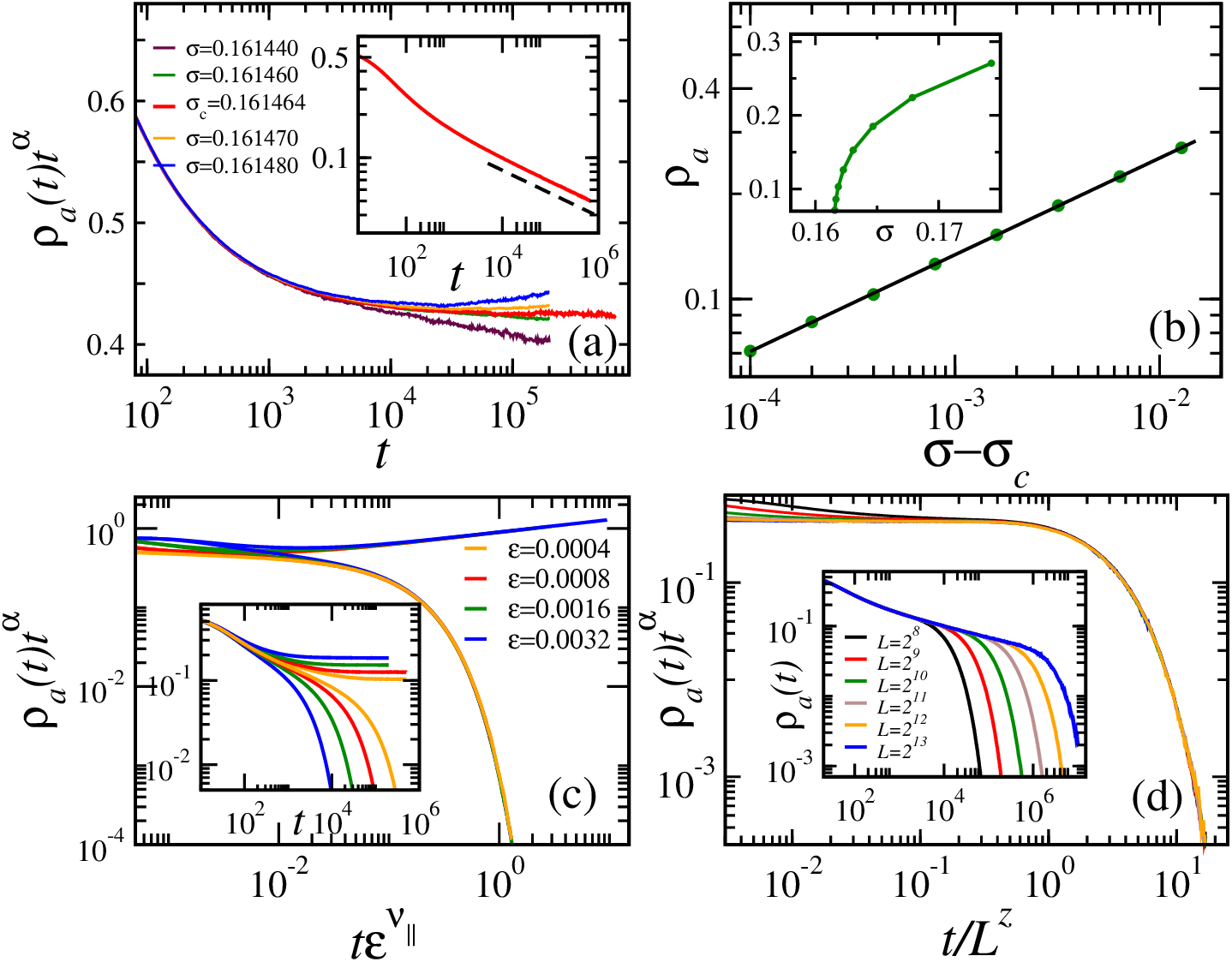}
 \caption{Critical behaviour for $\lambda=0.1:$ (a) determination of the critical point $\sigma_c$ and decay exponent $\alpha$ by plotting $\rho_a(t)t^\alpha$ for different values of $\sigma.$ Inset shows $\rho_a(t)$ versus $t$ at $\sigma_c$ in log-scale. (b) Log-scale  plot of the stationary $\rho_a$  against $\epsilon=\sigma-\sigma_c$  along with  the best fitted 
 line with slope $\beta=0.276(1).$ Inset shows the same in linear scale. 
(c) $\rho_a(t)t^\alpha$ as a function of the scaled variable $t\epsilon^{\nu_\shortparallel}$
for different values of $\epsilon.$ The upper and lower branch corresponds to super- and sub-critical regimes respectively.
(d) At  the critical point, $\rho_a(t)$   for different system sizes $L=2^8$-$2^{13}$  could be collapsed  following 
standard finite size  scaling   to obtain $z=1.58.$   The insets in (c) and (d) show the unscaled data.}
 \label{fig:lam0.1}
\end{figure}

We use the standard numerical techniques to study the critical behaviour of this driven dissipative Manna Model.  
The critical point $\sigma_c$ for a fixed $\lambda$ and the exponent $\alpha$ are estimated by measuring the temporal decay of density of active sites $\rho_a(t)$ starting from homogeneously active initial configurations for different values of $\sigma$ and looking for a power law decay. Alternatively, one can plot $\rho_a(t)t^\alpha$ versus $t$ for different values of $\sigma;$ the curve which remains constant in the large time limit corresponds to the critical point $\sigma_c.$ Fig. \ref{fig:lam0.1}(a) illustrates this procedure for $\lambda=0.1;$ resulting estimates are 
\bea
\sigma_c = 0.161464(2), \alpha = 0.16(1).
\eea

 The order parameter  exponent  $\beta,$  defined by $\rho_a \sim \epsilon^\beta$ where  $\epsilon=\sigma - \sigma_c,$ is estimated  
from the  log scale plot (Fig. \ref{fig:lam0.1}(b)) of $\rho_a$  versus $\epsilon,$ 
\bea
\beta = 0.276(1).
\eea

For $\sigma > \sigma_c$, the activity $\rho_a(t)$ obeys a scaling form 
$ \rho_a(t,\sigma) = t ^{-\alpha} {\cal F}\left( t(\sigma - \sigma_c)^{\nu_\shortparallel} \right).$
We estimate the exponent $\nu_\shortparallel$ by plotting $\rho_a(t)t^{\alpha}$ versus $t \epsilon^{\nu_\shortparallel}$ for different  $\epsilon$ (see Fig. \ref{fig:lam0.1}(c)) and looking for the best data collapse,
\bea
\nu_\shortparallel =1.73(2).
\eea
Finally we   study the finite size scaling. At the critical point $\sigma_c,$ the activity  in a finite system of size $L$ must decay as $\rho_a(t) = t^{- \alpha} {\cal G}(t/L^z).$   The dynamic exponent $z$ is estimated by plotting $\rho_a(t)t^{\alpha}$ against $t/L^z$ for different system sizes $L.$  This scaling collapse for $L=2^8-2^{13}$ is shown in  Fig. \ref{fig:lam0.1}(d) and yields
\bea
z =1.58(1).
\eea
All the critical exponents $(\alpha,$ $\beta,$  $\nu_\shortparallel$ and $z)$ estimated above are in very good agreement with DP and we can safely conclude that the critical behaviour of the driven dissipative  Manna Model for $\lambda=0.1$ is DP. 

\begin{figure*}[t]
  \centering \vspace*{-.5 cm}  \includegraphics[width=4.3 cm ]{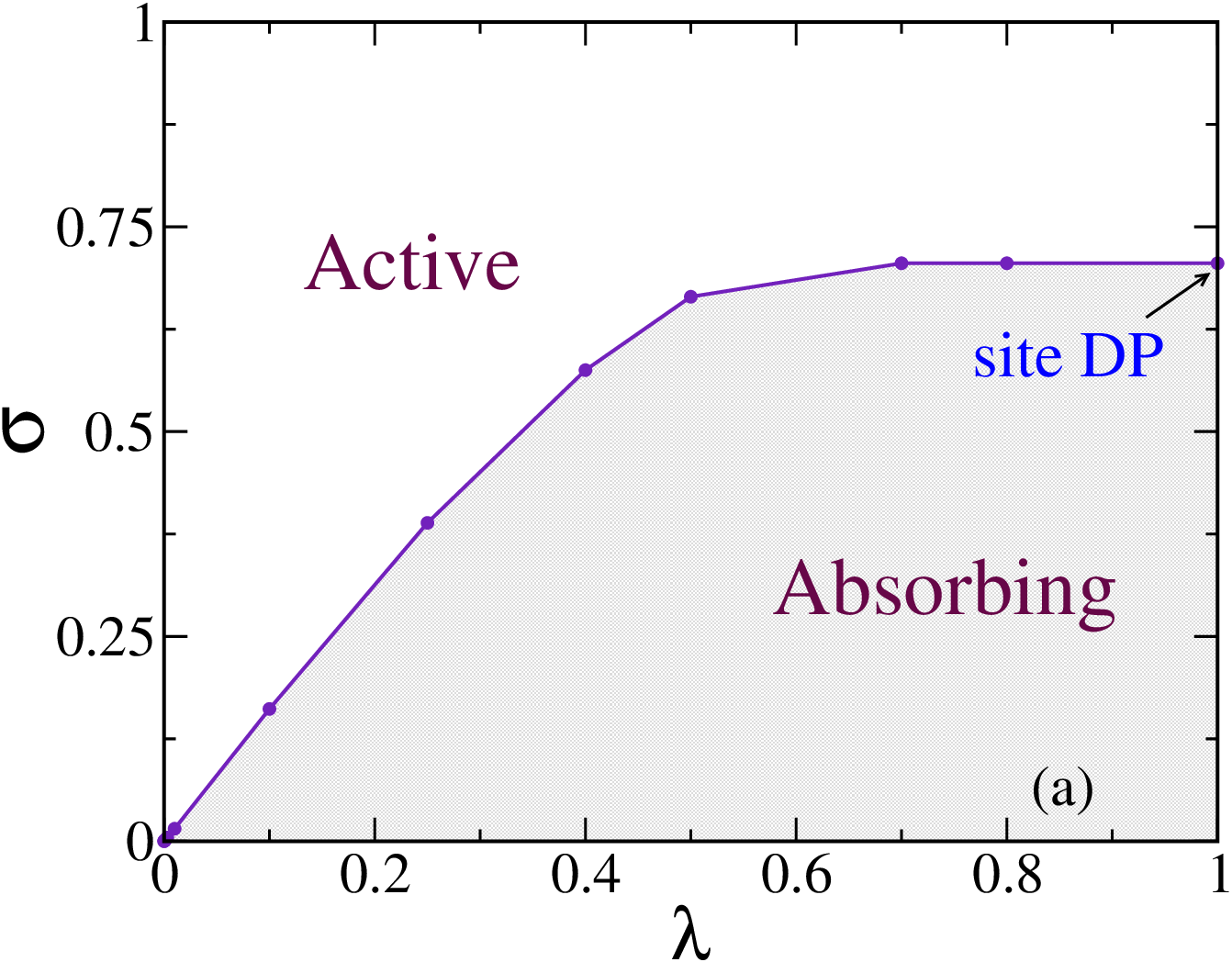}
 \includegraphics[width=8 cm ]{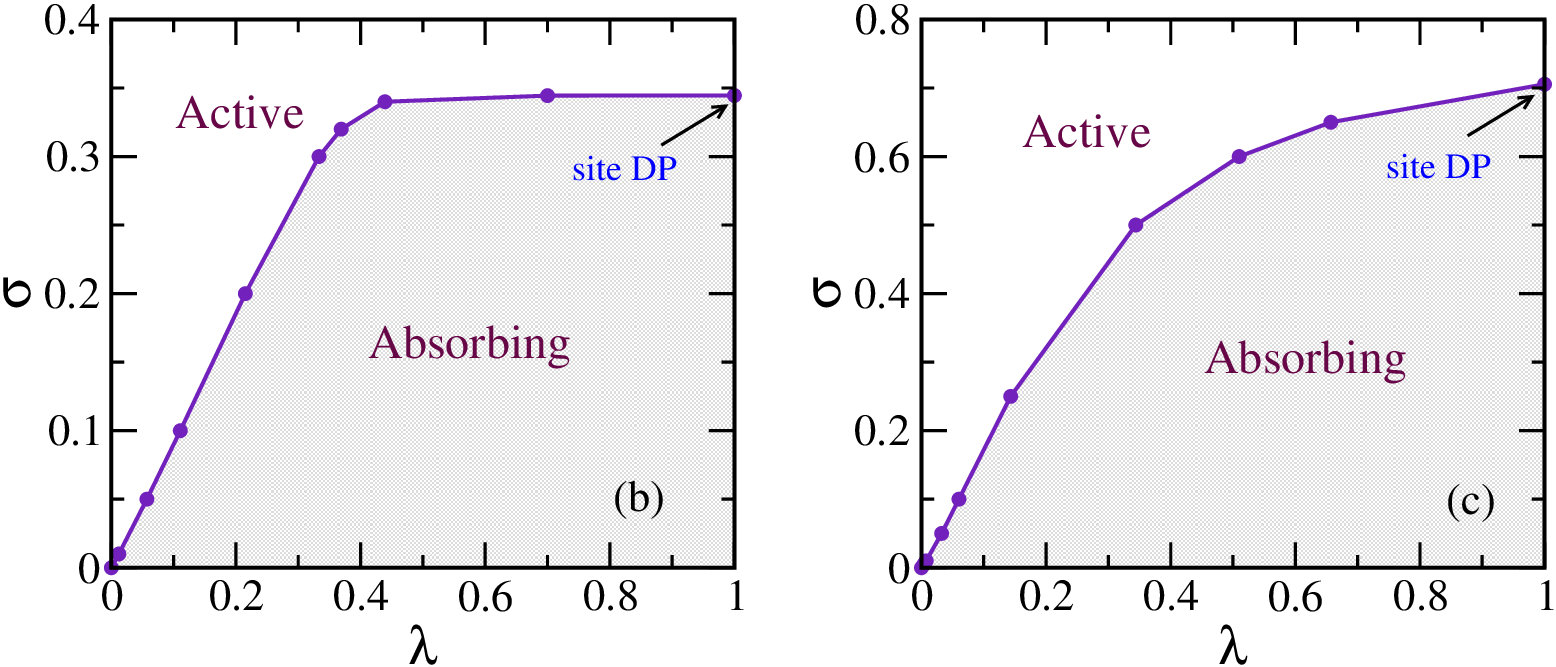}
 \caption {  Phase diagram in the $\lambda$-$\sigma$ plane : (a) continuous Manna Model in $1d$  (b) continuous Manna Model in $2d$
 (c) discrete Manna Model in $1d.$ The critical line separating absorbing and active phases end at the  site-DP critical points \cite{dp}. }\label{fig:phase}
\end{figure*}

\begin{table}
\begin{tabular}{l l l}
\hline 
$\lambda_c$ & $\sigma_c$   & $e_c$ \cr
\hline 
0.0002  &  0.0002796(2)   &   0.641\cr 
0.0004  &  0.000566(1) &  0.638 \cr 
0.001 &   0.001440(2) &  0.632  \cr 
0.003	& 0.004438(2) &  0.6238 \cr  
0.010 &   0.015312(2) &  0.606\cr 
0.100 &   0.161464(2)&    0.55  \cr  
0.25 &    0.38854(4) &   0.502\cr
0.4  &  0.57494(2) &   0.448\cr 
0.5  &    0.66460(2) &   0.385 \cr		
0.7	& 0.70548(1)  &  0.1782\cr
1 \cite{dp} &    0.705489   &    0 \cr
\hline
0 \cite{nomanna} &    0     &        0.65797  \cr
\hline
\end{tabular}
 \caption{Critical points for the driven dissipative continuous Manna Model.  {\clr Note that the average
 energy at the critical point $e(\lambda_c, \sigma_c),$ approaches the critical value $e_c = 0.65797$ obtained 
 in  the fixed-energy version \cite{nomanna}.} }\label{tab:1dccmm}
 \end{table}
There is nothing special about this particular value  $\lambda=0.1$ and one expects DP criticality  to continue 
for all $\lambda>0;$ we confirm  this   expectation  explicitly for a few other values of $\lambda,$ listed in 
Table \ref{tab:1dccmm}.  The phase diagram in $\lambda$-$\sigma$ plane   is  shown 
Fig. \ref{fig:phase} (a);  here the critical  line $\sigma_c(\lambda)$ separates the active phase 
from the absorbing one (shaded region).

{\clr  We have also studied the   critical   behaviour of the APT   taking a selected set of $0<\lambda<1,$ 
for  the  continuous model in $2d,$ and  for the discrete version in $1d.$   In all cases,    the criticalility 
is found to be in DP class - details are omitted since they are  only repetition  of  the procedure 
discussed for $d=1.$  Ciorresponding phase diagrams are   presented in Fig. \ref{fig:phase} (b) and (c) 
respectively.  Note that for  $d=2,$ the critical line starts at the site DP critical point $(\lambda_c=1, \sigma_c= 0.34457).$ }

{\clrb It is no surprise that}   all these models  show  DP transition The strength of  the arguments  primarily lies with  DP conjecture;   that 
a birth (given by Eq. (\ref{eq:III})), death ( Eq. (\ref{eq:I}) ) and diffusion ( Eq. (\ref{eq:II}) ) process  belongs to DP.  
The additional  feature,   finite drive acts  as a  multiplicative noise, which is already present in DP-Langevin equation \cite{dp}.

%
\section{Crossover from DP to SOC-like  scaling.}
A common feature of  all the   driven dissipative sandpile models models  discussed here is that   
they   undergo  an absorbing  phase transition in $\lambda$-$\sigma$ plane   with  a   line of critical  points   belonging to DP class.   This line   approaches  $\sigma \to 0$  for $\lambda\to 0.$  In this small  drive-dissipation   limit  the  scaling in the subcritical   region   splits, producing a crossover from  DP-scaling  close to critical line to SOC-like  scaling   when  one moves away.  
 Any   observable    $\la x \ra$   which  scales as $\Delta^{-a_x}$   in ordinary DP     crosses  over to 
 $\Delta^{-\tilde a_x}$   with  $\tilde a_x =a_x/\gamma$    resulting in multi-scaling, 
 \begin{equation}
   \la x \ra   = A_x     \Delta^{-a_x}  + B_x \Delta^{-\tilde a_x}, \label{eq:mult}
 \end{equation} 
 where non-universal coefficients  $A_x$   and    $B_x$  determine   the  crossover point. 
In the  main text  \cite{main_text} we have   seen that  both $\la s \ra$ and $\la T \ra$  obey 
this scaling form,   for   continuous    Manna   model in $d=1,2,$   and for discrete  Manna Model in $1d$ 
{\clr with  exponents  denoted   there  \cite{main_text} } as $a_s  \equiv \gamma $  and $a_T  \equiv  \tau.$ 

\begin{figure}
  \centering
 \includegraphics[width=8 cm ]{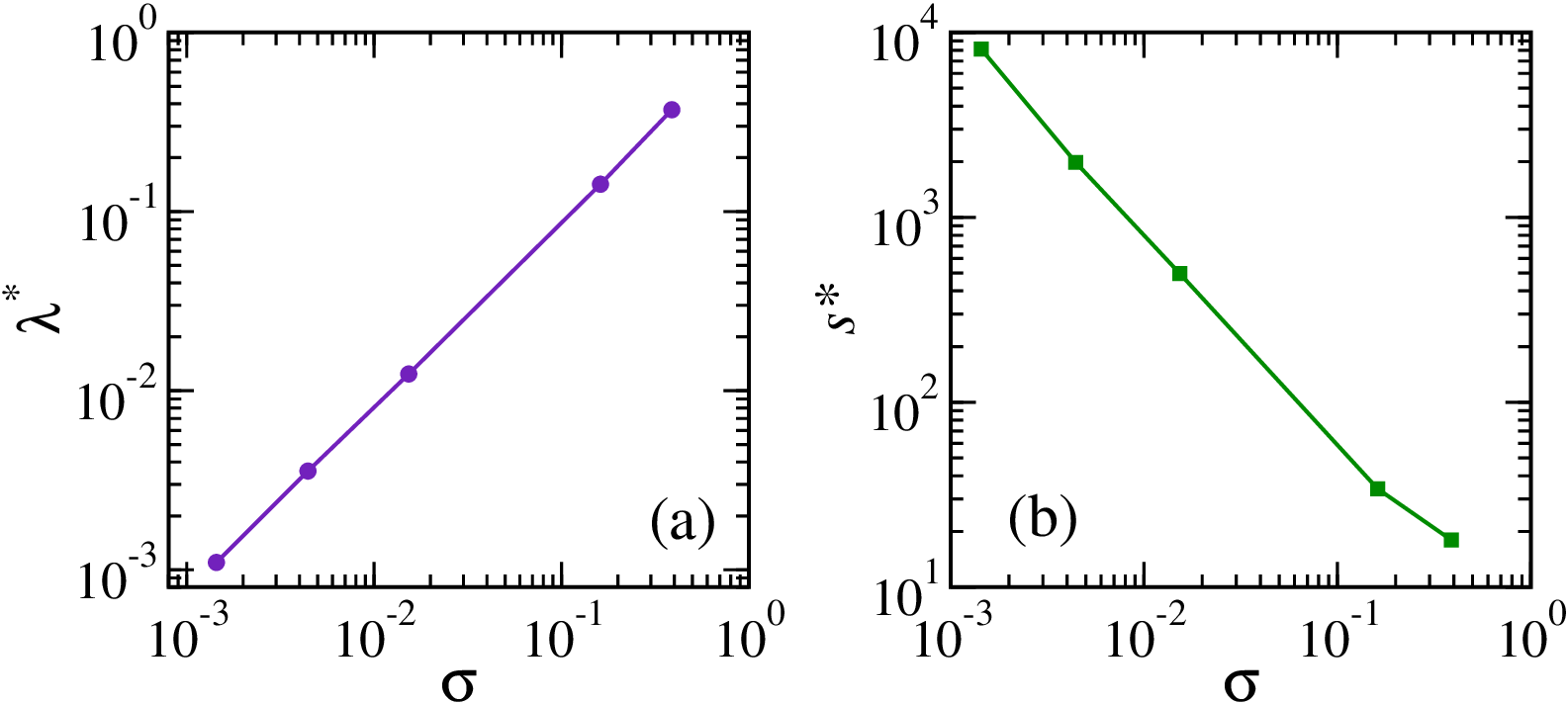}
 \caption  { For continuous Manna Model ($1d$) crossover occurs  near $\lambda=\lambda^*;$
 small clusters  ($\langle s\rangle < s^*$)   show SOC-like scaling whereas large ones   follow  DP.   
 (a) $\lambda^*$ vanishes  and  (b)  $s^*$   diverges  when  $\sigma \to 0.$
 The modified scaling behaviour takes over the whole scaling regime  in the SOC limit $\sigma = 0.$ 
 }\label{fig:crsover}
\end{figure}

For $\sigma=0,$ however, there is no crossover; these quantities $\la s \ra$ and $\la T \ra$ show a unique behaviour with the modified exponents $\tilde \gamma$ and  $\tilde \tau$, as can be seen from Fig. 5(c-d)  for  continuous model in $d=1,2$ and  from 
Fig. 4(b)  for 1d discrete model.  This provides an affirmative answer to a  question  of   utmost interest  
- whether  the  modified DP scaling   continues  to the $\sigma=0$ line, 
conventionally known  as the   bulk-dissipative  sand pile.  Let us  take  the example of 
$\la s \ra$  as a function of $\lambda$  in continuous   Manna Model   to  locate   the 
value  of $\lambda^*,$  where the crossover occurs  for a   given $\sigma.$    From   the best fit  of the 
data    to Eq.  \eqref{eq:mult}    we can approximately identify $\lambda^* = (B_s/A_s)^{1/(1-\gamma)}$  as the crossover point. 
In Fig. \ref{fig:crsover}(a) we have plotted  $\lambda^*$ as a function of  $\sigma$ in log scale; evidently   $\lambda^* \to 0$ 
as $\sigma \to 0.$ This indicates  that  the  ordinary DP scaling regime shrinks  and    finally disappears    when 
we approach the  SOC line $\sigma= 0.$    Thus, the   critical behaviour  at this SOC line is expected to  
be same as the `SOC-like' regime for $\sigma>0$ $i.e$ it is  DP with  modified exponents.   
This has been explicitly verified for  $d=1,2$  in main text \cite{main_text} . 

The crossover  in  $\la s \ra$     effectively 
means  that  small clusters with $\la s \ra< s^*$ follow  SOC-like scaling    and   only  the   large  
ones  follow  usual  DP scaling.   The   crossover scale $s^* = \la s \ra|_{ \lambda=\lambda^*}$  
depends on $\sigma$  and  as shown in Fig \ref{fig:crsover}(b),     $s^*$ diverges as $\sigma \to 0.$   
Thus,  for   $\sigma=0,$  the ordinary DP  feature is completely lost  and one observes only SOC-like scaling.